\documentclass[floatfix,showpacs,preprintnumbers,amsmath,amssymb,aps,twocolumn,prl,10pt]{revtex4}

\usepackage{amsfonts}
\usepackage[pdftex]{graphicx}
\usepackage{color}
\usepackage{bm}
\usepackage{multirow}

\newcommand{\bra}[1]{\langle #1|}
\newcommand{\ket}[1]{|#1\rangle}
\newcommand{\braket}[2]{\langle #1|#2\rangle}

\begin{document}
\title{Radio frequency spectra of Feshbach molecules in quasi-two dimensional geometries}
\author{Stefan K. Baur}
\affiliation{Cavendish Laboratory, J. J. Thomson Avenue, Cambridge CB3 0HE, United Kingdom}
\author{Bernd Fr\"ohlich}
\affiliation{Cavendish Laboratory, J. J. Thomson Avenue, Cambridge CB3 0HE, United Kingdom}
\author{Michael Feld}
\affiliation{Cavendish Laboratory, J. J. Thomson Avenue, Cambridge CB3 0HE, United Kingdom}
\author{Enrico Vogt}
\affiliation{Cavendish Laboratory, J. J. Thomson Avenue, Cambridge CB3 0HE, United Kingdom}
\author{Daniel Pertot}
\affiliation{Cavendish Laboratory, J. J. Thomson Avenue, Cambridge CB3 0HE, United Kingdom}
\author{Marco Koschorreck}
\affiliation{Cavendish Laboratory, J. J. Thomson Avenue, Cambridge CB3 0HE, United Kingdom}
\author{Michael K\"ohl}
\affiliation{Cavendish Laboratory, J. J. Thomson Avenue, Cambridge CB3 0HE, United Kingdom}
\date{\today}

\begin{abstract}
The line shape of radio frequency spectra of tightly bound Feshbach molecules in strong transverse confinement can be described by a simple analytic formula that includes final state interactions. By direct comparison to experimental data, we clarify the role of effective range corrections to two-body bound-state energies in lower dimensions.
\end{abstract}


\maketitle
Recent experimental progress in the creation of quasi-two dimensional (2D) atomic Fermi gases \cite{Martiyanov:2010kl,Frohlich:2011cr,Dyke:2011tg,Sommer:2011dq,Feld:2011nx,Zhang:ly} has provided important insights into the physics of fermionic pairing. Pairing phenomena in two dimensions \cite{Randeria:1989vn} are intriguing because of the presence of a confinement-induced bound state for a zero-range interaction potential of arbitrary strength and sign \cite{Petrov:2001fk}. This confinement-induced bound state significantly affects many-body pairing in the BEC-BCS crossover in two dimensions and leads to the appearance of a pairing pseudogap phase for weak pairing \cite{Randeria:1992zr, Trivedi:1995ys,Feld:2011nx}.

In the investigation of fermionic pairing, the single-particle excitation spectrum~\cite{Langmack:2011bs,Schmidt:2011ij, Ngampruetikorn:2011zr,Pietila:2012kh} probed by momentum-resolved~\cite{Feld:2011nx, Koehl2012} or momentum-integrated \cite{Frohlich:2011cr,Sommer:2011dq,Zhang:ly} radio-frequency (rf) spectroscopy plays a crucial role. Up to now, theoretical work on rf spectroscopy has focused on the strict two-dimensional case whereas experiments have been conducted in quasi-two dimensions. The latter refers to the situation in which tight confinement along one axis, usually with a harmonic frequency $\omega_z$, establishes a kinematically two-dimensional gas with $E_F,k_BT\ll \hbar \omega_z$. Here, $E_F$ is the Fermi energy of a two-dimensional Fermi gas, $k_B$ is Boltzmann's constant and $T$ is temperature. However, if the three-dimensional $s$-wave scattering length $a_s$, parameterizing the zero-range interaction potential, is smaller than the axial ground state $a_s<l_z=\sqrt{\hbar/m \omega_z}$, the relative wave function of the molecule explores higher transverse harmonic oscillator modes. Here, $m$ denotes the atomic mass.

In this Rapid Communication, we investigate the effects of quasi-2D confinement on rf spectra with a particular focus on the regime of strong pairing with $a_s>0$ and $E_B\gtrsim \hbar \omega_z$, both experimentally and theoretically. Specifically, we focus on three different aspects: first, the effect of finite range corrections, which are required to match experiment and theory in the limit of large binding energies, second the bound-bound transitions between confinement-induced states, and third the effects of final state interactions on rf spectra, which display significant differences between the strict-2D and the quasi-2D scenario. Our analytical theoretical results for binding energies and line shape functions compare well with the experimental data.

\begin{figure}
\includegraphics[width=\columnwidth]{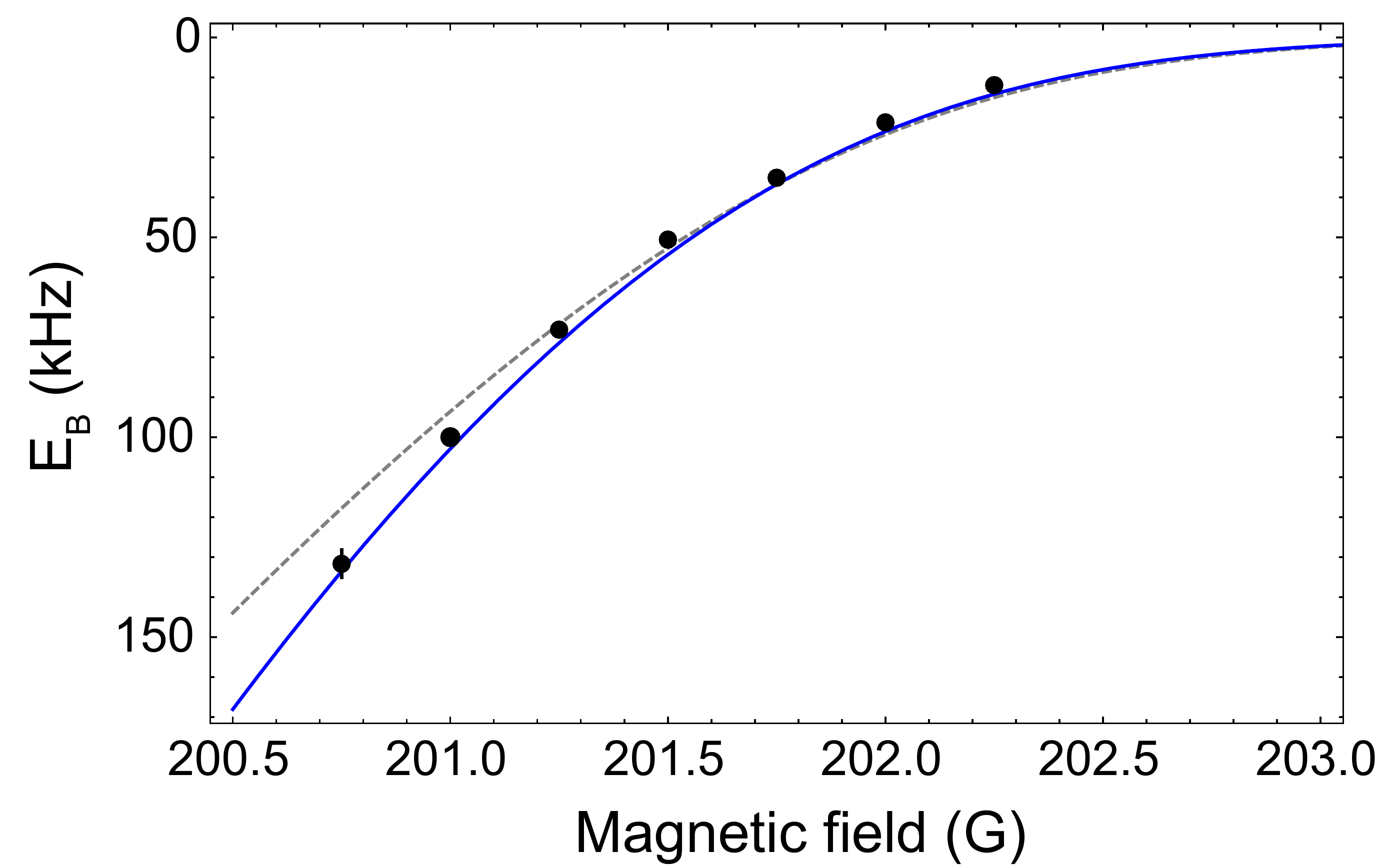}
\caption{(Color online) Binding energies of Feshbach molecules in a quasi-2D geometry. The bound-state energy including effective range corrections according to Eq. \eqref{eq:finiterangebs} is shown as the solid blue line. The theory for zero-range interaction only is the dashed gray line.}
\label{fig:bindingenergy}
\end{figure}

In our experiment, we prepare a 50/50 spin mixture of $^{40}$K atoms in the $|F=9/2,m_F=-9/2\rangle\equiv |1\rangle$ and $|F=9/2, m_F=-7/2\rangle\equiv |2\rangle$ states of the hyperfine ground state manifold~\cite{Frohlich:2011cr}. We achieve quantum degeneracy of two-dimensional Fermi gases in an optical lattice potential formed by a horizontally propagating, retro-reflected laser beam of wavelength $\lambda=1064$\,nm. The trapping frequency along the strongly confined direction is $\omega_z=2 \pi \times 75$\,kHz, which is calibrated by intensity modulation spectroscopy. The radial trapping frequency of the two-dimensional gases is $\omega_\perp=2\pi\times 127$\,Hz, and we confine on the order of $2\times 10^3$ atoms per two-dimensional gas at the center of the trap. Along the axial direction we populate approximately 30-40 layers of the optical lattice potential with an inhomogeneous peak density distribution. Starting from a weakly interacting gas ($a_s \sim 0$) at a magnetic field near 209\,G, we adiabatically increase the interaction strength by lowering the magnetic field at a rate of up to 0.25\,G/ms to a value near the Feshbach resonance at 202.1\,G. We apply an rf pulse with frequency $\nu_{\textrm{rf}}$ near 47\,MHz with a Gaussian amplitude envelope with a full width at half maximum of 230\,$\mu$s to transfer a fraction of atoms from the $|2\rangle$ state to the $|F=9/2,m_F=-5/2\rangle\equiv|3\rangle$ state. Atoms in the $|3\rangle$ state have a two-body s-wave scattering length of 250 Bohr radii with the $|1\rangle$ state. We turn off the optical lattice 100\,$\mu$s after the rf pulse, switch off the magnetic field and apply a magnetic field gradient to achieve spatial splitting of the three spin components in a Stern-Gerlach experiment during time-of-flight. Finally, we detect the atom numbers in each of the atomic states by absorption imaging.

\label{sec:boundstate}
We determine the binding energies of the molecules by recording the number of atoms transferred into the state  $|3\rangle$ and determine the threshold of the spectrum corrected for our spectral resolution of 1.5\,kHz \cite{Feld:2011nx}. In the following, we subtract the (trivial) contribution of the atomic Zeeman energy $E_Z$ from the rf frequency to obtain $\hbar \omega=E_Z-h \nu_\textrm{rf}$. In Figure \ref{fig:bindingenergy} we display the measured binding energies and compare with theoretical models. 
A pair of atoms interacting via zero range contact interactions under quasi-2D confinement has a binding energy given by the solution to the equation
\cite{Petrov:2001fk, Bloch:2008vn}
\begin{equation}
\label{eq:zerorangebs} g(E_B/\hbar \omega_z)=\frac{l_z}{a_s}
\end{equation}
where $\label{eq:integralrepforg} g(x)=\int_0^{\infty} \frac{du}{\sqrt{4 \pi u^3}} \left(1- e^{-x u}\left[ (1-e^{-2u})/(2u)\right]^{-1/2} \right)$  \cite{Bloch:2008vn}. This prediction is shown as the dashed line in Fig. \ref{fig:bindingenergy}. In the limit of large binding energy $E_B > \hbar \omega_z$ we find significant deviations from the quasi-2D binding energy. It has been argued that under tight confinement and for large binding energies, effective-range corrections to the standard zero-range model are of relevance~\cite{Blume:2002uq, Bolda:2002vn, Idziaszek:2005ve, Dickerscheid:2005zr,Naidon:2007ys, Zinner:2011oq}. These corrections can be viewed as an energy dependence of scattering length
\begin{eqnarray}
-\frac{1}{a_s(\epsilon)}=-\frac{1}{a_s}+\frac{k^2 r_{\text{eff}}}{2}+\ldots .
\end{eqnarray}
Here $\epsilon$ is the energy in the center of mass frame, $k=\sqrt{ m \epsilon}/\hbar$ and $r_{\text{eff}}$ is the effective range. By replacing $a_s \rightarrow a_s(\epsilon)$ with $\epsilon=\hbar \omega_z/2-E_B$, the effective range corrections to Eq. \eqref{eq:zerorangebs} take the form (see also the Supplemental Material)
\begin{eqnarray}
\label{eq:finiterangebs}
g(E_B/\hbar \omega_z)=\frac{l_z}{a_s}+\frac{r_{\text{eff}}}{2 l_z} \left(\frac{E_B}{\hbar \omega_z}-1/2 \right).
\end{eqnarray}
For open channel dominated Feshbach resonances, the effective range is given by~\cite{Flambaum:1999dz,Gao:1998fv}
\begin{eqnarray}
\label{eq:effectiverange}
r_{\text{eff}}=\frac{[\Gamma(1/4)]^4}{6 \pi^2} \bar{a} \left[ 1-2 \frac{\bar{a}}{a_s}+2 \left(\frac{\bar{a}}{a_s}\right)^2\right],
\end{eqnarray}
where $\bar{a}=2 \pi(C_6 m/\hbar^2)^{1/4}/[\Gamma(1/4)]^2$ is the mean scattering length. This implies that, in the vicinity of a Feshbach resonance, $^{40}$K atoms have an effective range of $r_{\text{eff}}\sim 10$nm (where we used $C_6=3897$ in atomic units)~\cite{Gribakin:1993vn,Chin:2010fk}. 
We plot the binding energy including effective range corrections as the solid line Fig.~\ref{fig:bindingenergy}. The excellent agreement with our measured binding energies demonstrates the importance of effective range corrections for quasi-2D confinement in the limit of tightly bound Feshbach molecules of the $|1\rangle$ and $|2\rangle$ states of $^{40}$K atoms. The agreement between and theory and experiment can presumably be further improved if the finite width of the  Feshbach resonance is taken into account~\cite{Bolda:2002vn,Julienne:2006kl}. This is expected to give a contribution of order $-2 R^*=-2 \bar{a}/s_{res} \sim -\bar{a}$ to the effective range \footnote{For the Feshbach resonance considered here, the width parameter is $s_{res}=2.2$~\cite{Chin:2010fk}.}.

\begin{figure*}
\includegraphics[width=\textwidth]{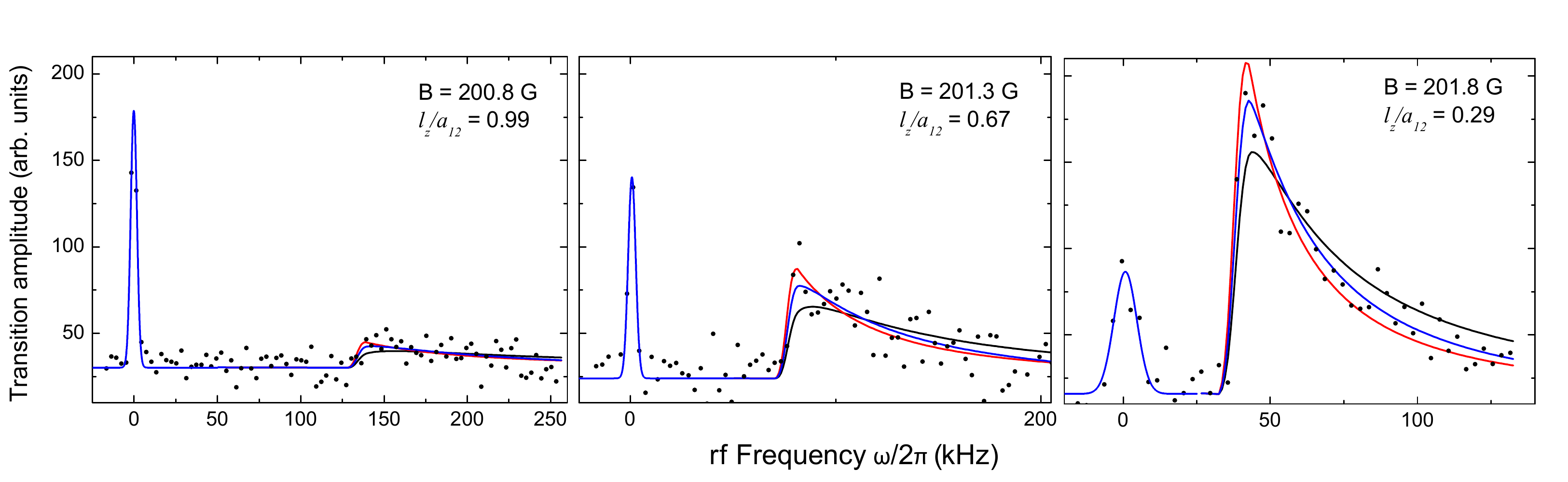}
\caption{Comparison of the experimental spectra in the molecular regime ($a_s > 0$) with the theoretical line shape functions described in the text for three different three-dimensional scattering lengths. We show theory curves for quasi-2D without final state interactions (red line), quasi-2D theory with final state interactions (blue line), and the strict 2D expression [see Eq. (13) of Ref.~\cite{Langmack:2011bs}] with 2D binding energies replaced with the quasi two-dimensional binding energies (black line). All curves were fitted with the same onset and vertical offset, and they are convoluted with a Gaussian of 2.5\,kHz width to account for the experimental resolution. Additionally, we have fitted a Gaussian of variable height and width centered at $\nu=0$ to account for the atomic peak. The absolute scale of $\Gamma(\nu)$ was least-squares fitted individually for each of the three theory curves.}
\label{fig:rfexperiment}
\end{figure*}

\label{sec:rf}
Next, we turn our attention to the effects of the quasi-2D nature on the rf spectra. First, we study the case without final state interactions. This situation is particularly simple as it contains only a bound-free transition. The rf transition is driven with a Rabi frequency $\Omega_R$, described by the operator
$\hat{V}=\hbar \Omega_R \left( \ket{2}\bra{3}+\ket{3}\bra{2} \right)$.
The transition rate is then given by Fermi's golden rule
\begin{eqnarray}
\label{eq:fermigoldenrule}
\Gamma_0(\omega)=\frac{2\pi}{\hbar} \sum_f |\bra{f}\hat{V}\ket{i}|^2 \delta(\hbar \omega+E_i-E_f),
\end{eqnarray}
which involves computation of the Franck-Condon overlaps ${\cal M}_{n}(q)\equiv \bra{f}\hat{V}\ket{i}/(\hbar \Omega_R)$ between the normalized bound-state relative wave function $\psi_B^{12}(\mathbf{r})=\left[ \pi l_z g'(E_B^{12}/\hbar \omega_z)\right]^{-1/2} \sum_{n} \phi_{n}(z) \phi_{n}(0) K_0(\kappa_{n} \rho)$, with $\kappa_{n}=\sqrt{m (E_B^{12}+\hbar \omega_z n)}/\hbar$, and free outgoing plane wave states $\phi^{13}_{n\mathbf{q}}(\mathbf{r})= e^{i \mathbf{q} \mathbf{r}}\phi_{n}(z)/2\pi$:
\begin{eqnarray}
{\cal M}_{n}^0(q)&=&\int d^3r \left[\phi_{n\mathbf{q}}^{13}(\mathbf{r})\right]^* \psi_B^{12}(\mathbf{r})\nonumber \\
&=&\frac{\hbar^2 \phi_{n}(0)}{m(E_{n\mathbf{q}}-E_B^{12})\sqrt{\pi l_z g'(E_B^{12}/\omega_z)}}.
\end{eqnarray}
Here, $\phi_{n}(z)$  is the wave function of the $n$-th transverse harmonic oscillator mode, $\mathbf{q}$ is the in-plane relative momentum and $E_{n\mathbf{q}}=\hbar^2 q^2/m+\hbar \omega_z n$. After summation over final states, we find the resulting line shape to be
\begin{eqnarray}
\Gamma_0(\omega) & = & 2 \pi \hbar \Omega_R^2 \sum_{n,\mathbf{q}} |{\cal M}^0_{n}(q)|^2 \delta(\hbar \omega-E_B^{12}-E_{n\mathbf{q}}) \nonumber \\&=& \frac{\sqrt{2\pi}\hbar \Omega_R^2}{m \omega^2 l_z^{2} g'(E_B^{12}/\hbar\omega_z)}\times \\
&&\hspace{11mm}\sum_{j=0}^{\infty} \frac{(2j-1)!!}{(2j)!!} \theta(\hbar \omega-E_B^{12}-2\hbar \omega_zj). \nonumber
\end{eqnarray}
Here, $g'(x)$ denotes the derivative of $g(x)$, $\theta(x)$ is the Heaviside function. For rf frequencies $\omega$ near the binding energy, the line shape shows a $\theta(\hbar \omega-E_B^{12})/\omega^2$ scaling which is found also in strict 2D \cite{Langmack:2011bs}. In this case, the rf spectrum takes the form
\begin{eqnarray}
\Gamma_0(\omega) &=& \frac{\Omega_R^2 C_{2D}}{2 m \omega^2}, \hspace{5mm} \textrm{for}\hspace{2mm} 0<\hbar\omega-E_B^{12}<2 \hbar \omega_z
\end{eqnarray}
in which we have introduced the two-dimensional contact parameter $C_{2D}=2 \sqrt{2\pi}/l_z^2 g'(E_B^{12}/\hbar \omega_z)$ \cite{Bertaina:2011fk, Langmack:2011bs}, which satisfies the 2D Tan relation. However, as the detuning from the binding energy becomes comparable to twice the axial oscillation frequency $2\omega_z$, additional local maxima appear in the spectrum, signaling the importance of higher axial modes. The tail of $\Gamma_0 (\omega)$ at very large frequencies $\hbar \omega \gg E_B, \hbar \omega_z$ behaves as
\begin{eqnarray}
\Gamma_0(\omega) \rightarrow \frac{2 \Omega_R^2}{\omega^{3/2} l_z g'(E_B^{12}/\hbar \omega_z)}=\frac{C_{3D}  \Omega_R^2}{2 \pi \omega^{3/2} (m/\hbar)^{1/2}}
\end{eqnarray}
using the three-dimensional contact~\cite{Braaten:2010kx} $C_{3D}=\frac{4 \pi m}{\hbar^2} \frac{d (-E_B^{12})}{d (-a_s^{-1})}$. This illustrates the fundamentally three-dimensional character of quasi-2D Feshbach molecules for short interatomic distances. Therefore, the strict-2D contact~\cite{Bertaina:2011fk, Langmack:2011bs} will only be revealed from rf spectra when confinement effects are negligible, i.e. when frequencies and binding energies much less than $\hbar \omega_z$ are considered.

Finally, we include final state effects. They have to be considered if the final state  $\ket{3}$ of the rf spin flip is still interacting with one of the initial states. In our experimental situation this will be the interaction between  $\ket{1}$ and  $\ket{3}$ which is parameterized by the scattering length $a_{13}$. The strict-2D problem of this scenario was studied by Langmack {\it et al}. \cite{Langmack:2011bs} and in a three dimensional setting by Chin and Julienne \cite{Chin:2005kx}. The effect of the final state interactions is two-fold: Firstly, a two-body confinement-induced bound-state appears and hence we expect the spectrum to  exhibit always a bound-bound transition in addition to the usual continuum~\cite{Bloch:2008vn}. Second, the line shape of the continuum spectrum changes as compared to the non-interacting case. Applying Eq. \ref{eq:fermigoldenrule} (Fermi's golden rule), we find that the rf spectrum of a $\ket{1}/\ket{2}$ bound-state takes the form
\begin{equation}
\Gamma(\omega)=2 \pi \hbar \Omega_R^2 |{\cal M}|^2 \delta(\hbar \omega-E_{B}^{12}+E_{B}^{13}) + \Gamma_c(\omega) 
\end{equation}
Here, ${\cal M}$ denotes the bound-bound transition amplitude and $E_{B}^{13}$ denotes the binding energy in the final state channel as a measure of the strength of the final state interactions. For the bound-bound transition amplitude, we obtain the analytic expression
\begin{eqnarray}
\label{eq:boundbound}
{\cal M}&=&\int d^3r \left[\psi_B^{13}(\mathbf{r})\right]^* \psi_B^{12}(\mathbf{r})\nonumber \\ &=&\frac{g(\xi_B^{12})-g(\xi_B^{13})}{(\xi_B^{12}-\xi_{B}^{13})\sqrt{g'(\xi_B^{12}) g'(\xi_{B}^{13})}}
\end{eqnarray}
by considering the overlap between the normalized bound-state wave functions $\psi_B^{ij}$,
where  $ij$ refers to the $\ket{1}/\ket{2}$ and $\ket{1}/\ket{3}$ channels (see also the Supplemental Material)~\cite{Chin:2005kx}.
Here, we introduced the dimensionless binding energies   $\xi_{B}^{ij}=E_{B}^{ij}/\hbar \omega_z$. Note that this result completely includes the effect of the confinement and in particular applies to molecules with a characteristic size of the order of the transverse confinement or smaller. Recently, this amplitude was also calculated numerically~\cite{Zhang:ly}.
The strict 2D expression
\begin{eqnarray}
\label{eq:boundbound2d}
{\cal M}_{2D}=\frac{\sqrt{E_B^{12} E_{B}^{13}}\log(E_B^{12}/E_{B}^{13})}{E_{B}^{12}-E_{B}^{13}}.
\end{eqnarray}
is obtained from Eq. \eqref{eq:boundbound} in the limit of small binding energies (see also ~\cite{Zhang:ly, Langmack:2011bs}). We find that Eq. \eqref{eq:boundbound2d} provides a reasonable estimate for the magnitude of the bound-bound transition amplitude even beyond the formal limit of its validity $E_B^{ij} \ll \hbar \omega_z$.

The transition rate into the continuum involves again the overlap between the $\ket{1}/\ket{2}$ bound-state and the unbound scattering states of the interacting final state channel. With the analytic expression of Ref. \cite{Petrov:2001fk} for these eigenstates $\psi_{n q}^{13}(\mathbf{r})$ 
in the presence of short range interactions between $\ket{1}/\ket{3}$ atoms, we obtain the expression (see also the Supplemental Material)
\begin{eqnarray}
{\cal M}_{n}(q)&=&\int d^3r\; \left[ \psi^{13}_{n q}(\mathbf{r}) \right]^* \psi^{12}_B(\mathbf{r})\; \nonumber \\
&=&\frac{\hbar^2 \phi_{n}(0) \left[ 1-  \frac{f^{13}(E_{n \mathbf{q}})}{f^{12}(E_{n \mathbf{q}})}\right]}{m(E_{n \mathbf{q}}-E_B^{12})\sqrt{\pi l_z g'(E_B^{12}/\omega_z)}}
\end{eqnarray}
which contains the scattering amplitude between atoms in states $\ket{i}$ and $\ket{j}$ at relative energy $E$~\cite{Petrov:2001fk}:
\begin{eqnarray}
f^{ij}(E)=\frac{2 \sqrt{2\pi}}{g(E_{B}^{ij}/\hbar \omega_z)-g(-E/\hbar \omega_z)}.
\end{eqnarray}
After summation over final states in Fermi's golden rule, the transition rate into the continuum takes the form
\begin{eqnarray}
\Gamma_c(\omega)=\Gamma_0(\omega) {\cal F}(\omega),
\end{eqnarray}
where effect of final state interactions on the continuum spectrum is described by the factor
\begin{eqnarray}
{\cal F}(\omega)=\left|1-\frac{f^{13}(\hbar \omega-E_B^{12})}{f^{12}(\hbar \omega-E_B^{12})}\right|^2.
\end{eqnarray}

When considering dissociation into the lowest axial mode at sufficiently low energies $E \ll \hbar \omega_z$, the scattering amplitudes $f^{ij}(E)$ may be approximated as
\begin{eqnarray}
f^{ij}(E)\approx \frac{4 \pi}{\sqrt{2 \pi}l_z/a_{ij}+\ln(-B \hbar \omega_z/(\pi E))},
\end{eqnarray}
where $B\approx 0.905$ ~\cite{Petrov:2001fk,Bloch:2008vn}. In this approximation, ${\cal F}(\omega)$ becomes
\begin{eqnarray}
\label{eq:lineshape}
{\cal F}(\omega)\approx \left| 1-\frac{\ln[\epsilon_{12}/(\hbar \omega-E_B^{12})]+i \pi}{\ln[\epsilon_{13}/(\hbar\omega-E_B^{12})]+i \pi} \right|^2
\end{eqnarray}
with $\epsilon_{ij}=(B/\pi) \hbar \omega_z e^{\sqrt{2\pi} l_z/a_{ij}}$. The energy scales $\epsilon_{ij}$ appearing in Eq. \eqref{eq:lineshape} are analogous to the scattering length $a_{ij}$ in three dimensions in the sense that they are the single parameter describing low energy collisions between atoms in the hyperfine states $\ket{i}$ and $\ket{j}$. However, note that $\epsilon_{ij}$ is different from the binding energy $E_B^{ij}$ for deeply bound molecules under quasi-2D confinement. We emphasize that Eq. \eqref{eq:lineshape} holds for any interaction strength at low energies (i.e. as long as $E \ll \hbar \omega_z$). This line shape is in general different from what would be obtained by using the strict two-dimensional formula of Refs. \cite{Zhang:ly, Langmack:2011bs} and replacing the 2D binding energies with the actual quasi two-dimensional binding energies $E_B^{12,13}$.

In Figure \ref{fig:rfexperiment} we display our measured rf spectra and compare with three different theoretical models: strict-2D and quasi-2D with no final state interactions, and quasi-2D with final state interactions. Generally, we find that the quasi-2D theories are more strongly peaked near the binding energy which appears to fit the data better. This difference in line-shape is particularly striking when the initial state is a tightly bound molecule or if the interactions in the final state are strongly repulsive (i.e. $a_{13} \gg l_z$). In contrast, for our experimental data for $^{40}$K with weak repulsive final state interactions ($l_z/a_{13}\approx 4.61\pm0.05$ over the range magnetic fields considered here), a quasi-2D line shape that does not take into account final state interactions still provides a good fit, whereas the strict-2D theory suggests a lesser-peaked line shape. The recent Duke experiment~\cite{Zhang:ly}, on the other hand, dissociates a weakly bound molecule into an outgoing channel with weak attractive interactions. In their case, $E_B^{12}$ ($E_{B}^{13}$) approximately coincide with $\epsilon_{12}$ ($\epsilon_{13}$), explaining why their strict-2D approach gives a good result for the line shape of a quasi-2D system.

In order to gain deeper insight into the more complicated and fundamentally important many-particle properties of confined quantum gases~\cite{Bertaina:2011fk, Parish:2011jl, Pietila:2012hc, Levinsen:2012kx}, a firm understanding of the interplay of confinement and few particle physics is crucial. In this Rapid Communication we provided analytic results for rf spectra of dilute quasi-2D paired gases in the two-body limit and compared those results to experimental data. Additionally, we demonstrated how effective range corrections contribute to the binding energy of confined quantum gases by explicitly comparing theory to experimental data. Beyond quasi-2D quantum gases, we believe our results can be generalized to quasi-one dimensional paired Fermi gases as experimentally studied in Refs.~\cite{Moritz:2005qf, Liao:2010bh}.

We gratefully acknowledge useful discussions with Nigel Cooper and Jesper Levinsen. The work has been supported by {EPSRC} through Grants No. EP/G029547/1 (M. K.) and No. EP/I010580/1 (S. B.), the Daimler-Benz Foundation (B.F.), Studienstiftung, and DAAD (M.F.).

\clearpage
\newpage
\newpage
\section*{Supplemental Material for Radio frequency spectra of Feshbach molecules in quasi-two dimensional geometries}
\section*{Binding energy and effective range corrections under confinement}
Theoretically, two interacting atoms in hyperfine states $\ket{1}$ and $\ket{2}$ under quasi-2D harmonic confinement are described by the Hamiltonian
\begin{eqnarray}
H=\sum_{i=1,2}\left[-\frac{\hbar^2 \partial_{\mathbf{r}_i}^2}{2m}+V_H(z_i)\right]+V_{int}(\mathbf{r}_1-\mathbf{r}_2),
\end{eqnarray}
where the interaction $V_{int} (\mathbf{r})$ is the interaction potential and $V_H(z)=\frac{1}{2} m \omega_z^2 z^2$. After separating into the relative coordinate $\mathbf{r}=\mathbf{r}_1-\mathbf{r}_2$ and the center of mass coordinate $\mathbf{R}=(\mathbf{r}_1+\mathbf{r}_2)/2$ one has
\begin{eqnarray}
H&=&H_{rel}+H_{CM} \\
H_{rel}&=&-\frac{\hbar^2 \partial_{\mathbf{r}}^2}{2 \mu}+\frac{1}{2}\mu \omega_z^2 z^2+V_{int}(\mathbf{r})\\
H_{CM}&=&-\frac{\hbar^2 \partial_{\mathbf{R}}^2}{2 M}+\frac{1}{2} M \omega_z^2 Z^2
\end{eqnarray}
with reduced mass $\mu=m/2$ and total mass $M=2m$. For a short ranged potential $V_{int}(\mathbf{r})$ with range $r_0 \ll l_z$, we can find an approximate solution to the two-body problem by first noting that the ansatz~\cite{Idziaszek:2005ve}
\begin{eqnarray}
\psi_E(\mathbf{r}) \propto G_E(\mathbf{r},0)
\end{eqnarray}
where $G_E(\mathbf{r}_1,\mathbf{r}_2)$ is the Green's function of the Hamiltonian $H_{rel}$. $G_E(\mathbf{r},0)$ satisfies the Schr\"odinger equation for the relative coordinate Hamiltonian away from the origin $\mathbf{r}=0$. Close to the origin ($r_0<r\ll l_z$) we match this outer solution to logarithmic derivative of the asymptotic free space scattering solution of the interaction potential $V_{int}(\mathbf{r})$
\begin{eqnarray}
k_{\bar{E}} \cot(\delta_{\bar{E}})=\left. \frac{\partial_r(r \psi_{\bar{E}}(\mathbf{r}))}{r \psi_{\bar{E}}(\mathbf{r})} \right|_{r=0},
\end{eqnarray}
where $\bar{E}\equiv E+\hbar \omega_z/2$, $k_{\bar{E}}=\sqrt{2 \mu \bar{E}}/\hbar$ and $\delta_{\bar{E}}$ is the s-wave phase shift at energy $\bar{E}$. With the usual effective range expansion $k \cot(\delta)=-1/a_s+k^2 r_{\text{eff}}/2+ \ldots$ and the expansion of $G_E(\mathbf{r},0)$ from Refs. \cite{Petrov:2001fk2,Bloch:2008vn}
\begin{eqnarray}
G_E(\mathbf{r},0) \sim \frac{1}{r}-\frac{g(-E/\hbar \omega_z)}{l_z},
\end{eqnarray}
we arrive at eq. 3 of the main text.

\section*{Effect of confinement and final state interactions on radio frequency spectra in quasi-2D}
\subsection*{Simple considerations: 2D Limit}

It is instructive to study the problem of molecule dissociation in a strict two-dimensional setting~\cite{Langmack:2011bs2, Zhang:ly2}. Such a description should be valid whenever the binding energy $E_B^{12}$ ($E_{B}^{13}$) of the molecule in initial (final) state channels and the energy of the RF-photon is much less than the spacing of transverse harmonic oscillator modes $\omega_z$. In general the RF spectrum $\Gamma(\omega)$ of molecules can be calculated from Fermi's Golden rule~\cite{Chin:2005kx2}
\begin{eqnarray}
\Gamma^{2D}(\omega)=2 \pi \sum_{f} |\braket{\psi_B}{\psi_f}|^2 \delta(\omega-E_B^{12}-E_f)
\end{eqnarray}
The summation over final states includes a bound-state at energy $E_f=-E_{B}^{13}$ and a continuum of unbound atoms with  $E_f=\mathbf{k}^2$ and relative momentum $\mathbf{k}$. For brevity, we use units where $\hbar=m=\Omega_R=1$. The matrix element for the bound-bound transition amplitude ${\cal M}_{2D}$ is given by the overlap integral
\begin{eqnarray}
{\cal M}_{2D}=\int d^2r\; \left[ \psi^{12}_B(\mathbf{r}) \right]^* \psi_{B}^{13}(\mathbf{r})
\end{eqnarray}
where $\psi_B^{12}(\mathbf{r})$ [$\psi_{B}^{13}(\mathbf{r})$] is the molecule wavefunction in the initial [final] state, respectively. At bound-state energy $E_B^{ij}=\kappa_{ij}^2$, the normalized wavefunction for short-range attractive s-wave interactions in 2D is
\begin{eqnarray}
\psi_{B}^{ij}(\mathbf{r})=\kappa_{ij} K_0(\kappa_{ij} \rho)/\sqrt{\pi}
\end{eqnarray}
which gives
\begin{eqnarray}
\label{eq:boundbound2d}
{\cal M}_{2D}=\frac{\sqrt{E_B^{12} E_{B}^{13}}\log(E_B^{12}/E_{B}^{13})}{E_{B}^{12}-E_{B}^{13}}.
\end{eqnarray}
We note that this result makes sense from a qualitative point of view: When interactions in final and initial state are equal, one has ${\cal M}_{2D}=1$ (as all weight is concentrated in the bound-bound transition), whereas for vanishing final state interactions (i.e. $E_{B}^{13}=0$ or $E_{B}^{13}=\infty$), ${\cal M}_{2D}$ tends to zero.

The bound-free continuum can be calculated from the matrix element between a s-wave scattering state $\ket{\psi^{(+)}_k}$ and $\ket{\psi_B}$. Such a state satisfies a Lippmann-Schwinger equation
\begin{eqnarray}
\ket{\psi^{(+)}_k}=\ket{\phi^{(0)}_k}+\frac{1}{k^2-H_0+i \eta} V' \ket{\psi^{(+)}_k}
\end{eqnarray}
and is guaranteed to have the same normalization as the solutions to the free Hamiltonian $\ket{\phi^{(0)}_k}$. For a 2D short range potential the Lippmann-Schwinger equation is solved in the s-wave channel by~\cite{landaulifshitz, Petrov:2001fk2}
\begin{eqnarray}
\psi^{(+)}_k(\rho)=\frac{J_0(k \rho)}{2\pi}- \frac{1}{2\pi}\frac{i}{4} f^{13}(k^2) H_0^{(1)}(k \rho)
\end{eqnarray}
with the 2D scattering amplitude for the final state channel
\begin{eqnarray}
f^{13}(E)=\frac{4 \pi}{i\pi+\log(E_B^{13}/E)}.
\end{eqnarray}
Here we picked the normalization of the scattering states such that
\begin{eqnarray}
\int d^2r\; \left[\psi^{(+)}_{k}(\mathbf{r})\right]^*\psi^{(+)}_{k'}(\mathbf{r})=\frac{1}{2 \pi k} \delta(k-k')
\end{eqnarray}
In this normalization, the bound-free transition amplitude ${\cal M}_{2D}(k)$ becomes
\begin{eqnarray}
{\cal M}_{2D}(k)&=&\int d^2r\; [\psi_B^{12}(\mathbf{r})]^* \psi_k^{(+)}(\mathbf{r}) \nonumber\\&=&\frac{\sqrt{E_B^{12}/\pi}}{k^2+E_B^{12}}\left[1-\frac{f^{13}(E)}{f^{12}(E)}\right].
\end{eqnarray}
Evaluating the sum over final momentum states we obtain a result for the bound-free spectrum part of the spectrum $\Gamma_c^{2D}(\omega)$:
\begin{eqnarray}
\Gamma_c^{2D}(\omega)&=&2 \pi \int d^2k\; |{\cal M}_{2D}(k)|^2 \delta(\omega-E_B^{12}-k^2)\nonumber\\
&=& \Gamma_0^{2D}(\omega) {\cal F}^{2D} (\omega)
\end{eqnarray}
where $\Gamma_0^{2D}(\omega)=\theta(\omega-E_B^{12})\frac{ 2\pi E_B^{12} }{\omega^2}$ is the RF-spectrum without final state interactions and
\begin{eqnarray}
\label{eq:2dlineshape}
{\cal F}^{2D}(\omega)=\left| 1-\frac{f^{13}(\omega-E_B^{12})}{f^{12}(\omega-E_B^{12})} \right|^2
\end{eqnarray} is a correction term accounting for final state interactions. Note that the complete spectrum
\begin{eqnarray*}
\Gamma^{2D}(\omega)=|{\cal M}_{2D}|^2 \delta(E_{B}^{13}-E_{B}^{12}+\omega)+\Gamma_0^{2D}(\omega) {\cal F}^{2D}(\omega)
\end{eqnarray*}
satisfies the sum-rule
\begin{eqnarray}
\int_{-\infty}^{\infty} d\omega\; \Gamma^{2D}(\omega)=2\pi.
\end{eqnarray}
\subsection*{Effect of closed channels}
We will now turn our attention to the problem of a three-dimensional gas confined into two dimensions by tight one-dimensional harmonic confinement. In such a potential the relative and center of mass coordinates separate, so we only need to consider the Hamiltonian for the relative coordinate.  In order to calculate the bound-bound and bound-free transition rates, we first need to know the normalized wavefunction of the two-body bound state in this geometry. The Green's function at energy $E$ in the harmonic oscillator potential is given by~\cite{Petrov:2001fk2}
\begin{eqnarray*}
G_{E}(\mathbf{r},0)=\sum_{\nu} \phi_{\nu}(z) \phi_{\nu}(0) \left\{ \begin{array}{c} (i/4) H_0^{(1)}(q_{\nu} \rho); \hspace {3mm} q_{\nu}^2>0 \\ K_0(|q_{\nu}| \rho)/(2\pi); \hspace {3mm} q_{\nu}^2<0 \end{array}\right.,
\end{eqnarray*}
where $q_{\nu}^2=E-\nu$ and the transverse harmonic oscillator states have $\phi_{\nu}(0)=(\nu-1)!!/[\sqrt{\nu!} (2\pi l_z^2)^{1/4}]$. We can use the explicit expression for the bound-state wave function $\psi_B^{ij}(\mathbf{r})={\cal N}_{ij} G_{-E_B^{ij}}(\mathbf{r})$ in terms of $G_E(\mathbf{r},0)$ to obtain the normalization factor 
\begin{eqnarray}
{\cal N}_{ij}=\sqrt{\frac{4\pi}{l_z g'(E_B^{ij}/\omega_z)}}
\end{eqnarray}
where $E_B^{ij}$ is the two-body binding energy, satisfying $g(E_B^{ij}/\hbar \omega_z)=l_z/a_{ij}$. The normalized wavefunction of the two-body bound-state is thus given by
\begin{eqnarray*}
\psi_B^{ij}(\mathbf{r})=\frac{1}{\sqrt{\pi l_z g'(E_B^{ij}/\omega_z)}} \sum_{\nu} \phi_{\nu}(z) \phi_{\nu}(0) K_0(|q_{\nu}| \rho).
\end{eqnarray*}
With this wavefunction it is straightforward to calculate the bound-bound transition amplitude $\cal{M}$
\begin{eqnarray}
{\cal M}&=&\frac{1}{l_z \sqrt{g'(\xi_B^{12}) g'(\xi_{B}^{13})}} \sum_{\nu} |\phi_{\nu}(0)|^2 \frac{\log \left( |q_{\nu}|^2/|q'_{\nu}|^2\right)}{|q_{\nu}|^2-|q'_{\nu}|^2}\nonumber\\ &=&\frac{g(\xi_B^{12})-g(\xi_{B}^{13})}{(\xi_B^{12}-\xi_{B}^{13})\sqrt{g'(\xi_B^{12}) g'(\xi_{B}^{13})}},\label{eq:boundboundq2d}
\end{eqnarray}
where $\xi_B^{ij}=E_B^{ij}/\hbar \omega_z$.
We note that for equal interactions in initial and final states (i.e. $\xi_B^{12}=\xi_{B}^{13}$), ${\cal M}=1$. Also note that for small $\xi_B^{12}$, $\xi_{B}^{13} \ll 1$, eq. \eqref{eq:boundboundq2d} reduces to the 2D result eq. \eqref{eq:boundbound2d}. Analogous to the simpler 2D limit we can obtain the contribution of the bound-free continuum from the overlap between scattering state and bound state wavefunction. For the final scattering states we use the basis of retarded scattering states given described in Ref. \cite{Petrov:2001fk2}
\begin{eqnarray}
\psi^{(+)}_{\nu,\mathbf{q}}(\mathbf{r})=\frac{1}{2\pi}\phi_{\nu}(z) J_0(q \rho)+\frac{A_{\nu}'}{2\pi} G_{E}(\mathbf{r},0)
\end{eqnarray}
For the precise expressions of the coefficients $A_{\nu}'$ see Ref.~\cite{Petrov:2001fk2}. Here the prime indicates that we consider the final state channel. By parity symmetry, the overlap matrix element ${\cal M}_{\nu}(q)$ vanishes for odd $\nu$. For even $\nu$ one obtains
\begin{eqnarray*}
{\cal M}_{\nu}(q)&=&\int d^3r\; \psi^{12}_B(\mathbf{r})\; \psi^{(+)}_{\nu,q}(\mathbf{r})\\
&=&\left. \frac{\phi_{\nu}(0) \left[ 1-  \frac{f^{13}(E)}{f^{12}(E)}\right]}{(E-E_B^{12})\sqrt{\pi l_z g'(E_B^{12}/\omega_z)}} \right|_{E=q^2+\nu}.
\end{eqnarray*}
The transition rate into the continuum is given by
\begin{eqnarray*}
\Gamma_c(\omega)&=& 2 \pi \sum_{\nu,q} |{\cal M}_{\nu}(q)|^2 \delta(\omega-E_B^{12}-q^2-\nu \omega_z)\\&=&2\pi^2\sum_{\nu} |{\cal M}(\sqrt{\omega-\nu \omega_z-E_B^{12}})|^2\\&=&\left[ \frac{\sqrt{2 \pi}}{\omega^2 l_z^2 g'(\xi_B) } \sum_{j=0}^{\infty}\frac{(2j-1)!!}{(2j)!!} \theta(\omega-E_B^{12}-2j)\right]\\ &&\times \left |1-\frac{f^{13}(\omega-E_B^{12})}{f^{12}(\omega-E_B^{12})}\right|^2.
\end{eqnarray*}
Note that this result is again of the form $\Gamma_c(\omega)=\Gamma_0(\omega) {\cal F}(\omega)$, where $\Gamma_0(\omega)$ is spectrum without final state interactions and the factor
\begin{eqnarray}
{\cal F}(\omega)=\left|1-\frac{f^{13}(\omega-E_B^{12})}{f^{12}(\omega-E_B^{12})}\right|^2.
\end{eqnarray}
We finally note that it is convenient to use a closed form expression for the sum
\begin{eqnarray}
\sum_{j=0}^{\infty} \frac{(2j-1)!!}{(2j)!!} \theta(x-2j)=\frac{2}{\sqrt{\pi}} \frac{\Gamma(\lfloor x/2 \rfloor+3/2)}{\lfloor x/2 \rfloor !}
\end{eqnarray}
as given in Ref. \cite{Petrov:2001fk2}. With this expression one can directly confirm that the tails of RF-spectrum without final state interactions obeys the 3D Tan relations.
\section*{Numerical evaluation of $g(\epsilon)$}
Since the expression for $g(\epsilon)$ via the summation of Ref. \cite{Petrov:2001fk2} converges very slowly we will give a numerically more efficient scheme to evaluate differences of the form
\begin{eqnarray}
W(\epsilon_1,\epsilon_2)=g(\epsilon_1)-g(-\epsilon_2)
\end{eqnarray}
where $\epsilon_1,\epsilon_2>0$. Using the integral representation for $g(\epsilon)$ of Ref. ~\cite{Bloch:2008vn} one obtains
\begin{eqnarray}
g'(\epsilon)=\frac{\Gamma \left(\frac{\epsilon }{2}\right)}{2 \sqrt{2} \Gamma \left(\frac{\epsilon
   +1}{2}\right)}.
\end{eqnarray}
after exchanging integration and differentiation. We can then numerically compute
\begin{eqnarray}
W(\epsilon_1,\epsilon_2)=\int_C g'(z) dz,
\end{eqnarray}
where $C$ is a contour in the lower complex halfplane (e.g. we used straight lines along $\epsilon_1 \rightarrow -i \rightarrow -\epsilon_2$).

\end{document}